\documentstyle[epsf,editedvolume]{crckapb} 
\def\plottwo#1#2{\centering \leavevmode
\epsfxsize=.5\textwidth \epsfbox{#1} \hfil
\epsfxsize=.5\textwidth \epsfbox{#2}}
\def\plottwob#1#2{\centering \leavevmode
\epsfxsize=.52\textwidth \epsfbox{#1}
\epsfxsize=0.95\textwidth \epsfbox [120 215 750 380] {#2}}
\def\plotfiddle#1#2#3#4#5#6#7{\centering \leavevmode
\vbox to#2{\rule{0pt}{#2}}
\includegraphics{#1}}

\newcommand{\etal}{{\it et al.}\ }

\begin{opening}
\title{EVOLUTIONARY POPULATION SYNTHESIS MODELS OF\protect\\
PRIMEVAL GALAXIES: A CRITICAL APPRAISAL}
\subtitle{ }

\author{A. BUZZONI}
\institute{Osservatorio Astronomico di Brera\\
           Via Bianchi, 46 23807 - Merate (Lc), Italy}

\end{opening}

\runningtitle{SYNTHESIS MODELS OF PRIMEVAL GALAXIES}

\begin{document}

\section{Introduction}

A recognized problem when searching for primeval galaxies at cosmological
distances is a definition of a firm selection criterion (alternative to a 
direct but extremely time-demanding measure of $z$) to single out 
high-redshift candidates from the {\it plethora} of other more or 
less peculiar objects (i.e. AGN, starburst galaxies etc.) at lower distances.

The case of the HST ``Deep Field'' (HDF) observations (Williams \etal 1996) is 
especially relevant in this regard as galaxies up to $z = 3.2$ have been 
detected in the field (Steidel \etal 1996).

Taking advantage of the selective effect on galaxy apparent colors due
to foreground $L_\alpha$ absorption from intergalactic Hydrogen clouds 
(the so-called $L_\alpha$ forest), Madau \etal (1996) have been able to select 
star-forming galaxies in the HDF up to a fiducial redshift of $z \sim 4.5$.

This method complements the more direct empirical approach 
of {Lanzet-ta} \etal (1996) who probed galaxy redshift by fitting observed colors 
with spectra of local templates of appropriate morphological 
type. The latter procedure is however prone to a subtle internal inconsistency 
since it implicitly neglects any evolutionary effect.

On the other hand, the main drawback of Madau's \etal method
resides in its selection bias toward a special class of candidates
(the so-called ``dropout'' objects). Clearly, this makes difficult
to assess how representative is the selected subsample of the whole
galaxy population at the relevant redshift.

In this framework, a theoretical approach relying on evolutionary population 
synthesis models could help refining the search criteria in deep galaxy surveys 
on the basis of a better knowledge of the expected apparent photometric 
properties of high-redshift objects.
The following is a brief discussion reviewing some relevant aspects of
the question in order to allow a more critical appraisal to primeval 
galaxy recognition.

\section{Global Energetics in Model Stellar Populations}

It is especially convenient to consider a model galaxy
in terms of its composing simple stellar populations (SSPs),
that is of single generations of coeval stars with fixed distinctive 
parameters like metallicity, initial mass function (IMF) etc.

A basic problem when tracking SSP evolution deals with a correct 
evaluation of Main Sequence (MS) and Post-Main Sequence relative contribution 
to the SSP total luminosity. 
From the theoretical and operational point of view in fact these two 
main building blocks are managed in a substantially different way.

While Post-MS star distribution is simply proportional to the lifetime 
duration in each isochrone bin (so that $n_i/n_j = t_i/t_j,~~\forall i, j$),
MS number counts are on the contrary modulated by the IMF
(namely $n_i \propto M_i^{-s}$ for a standard Salpeter power law).
We need therefore to know how to scale Post-MS {\it as a whole} with respect 
to the MS component in the model SSP.

Clearly, a simple match by grafting contiguous isochrone bins 
at the MS top and at the Post-MS bottom  {\it cannot be a safe solution} 
in this regard as any  uncertainty and numerical ``noise'' in the evaluation 
of both extrema would accordingly reverberate into a 
magnified scatter in the contribution of each SSP building block.

Rather than relying on such a {\it differential} normalization procedure, 
one could take advantage of an {\it integral} approach 
featuring SSP global properties. This can be done via the so-called 
``Fuel Consumption Theorem'' (FCT) (Renzini and Buzzoni 1986) dealing 
with SSP energetics.
Starting from the general relation in Renzini and Buzzoni (1986, their eq. 14), 
it is possible to state the theorem as follows:

$$ {{L_{PMS}}\over{L_{tot}}} = {\cal B} \times {\rm Fuel} = (1.76 \pm 0.4)~m_H. \eqno (1) $$

{\it This simply relates SSP Post-MS luminosity to the total amount of fuel 
spent in Post-MS evolution by stars of mass $M_{TO}$ leaving the MS Turn Off point.}

In the equation, the normalization factor $\cal B$ is the so-called ``specific 
evolutionary flux''; it turns to be about ${\cal B} = 1.7\pm 0.4~10^{-11}$ [L$_\odot^{-1}$ yr$^{-1}$]
(Buzzoni 1989). The term $m_H$ is the exhausted fuel expressed in Hydrogen-equivalent solar masses.
This quantity is a straightforward output of stellar evolution theory.

It might be interesting to check whether FCT prescriptions are consistently
met by current theoretical synthesis codes, some of them extensively 
adopted for cosmological studies. Following the original discussion in 
Buzzoni (1995), in Fig. 1 we have compared template models for 
present-day elliptical galaxies according to the different theoretical sources.

\begin{figure}
\plotfiddle{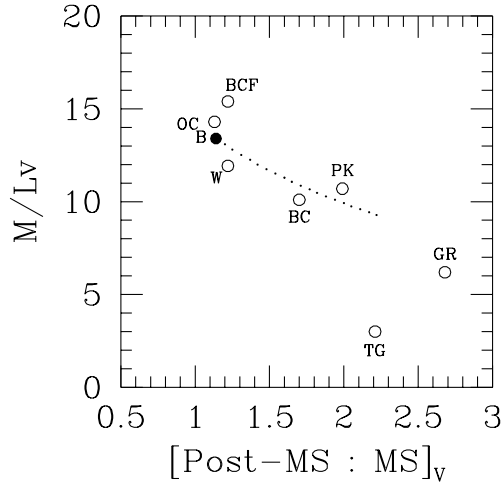}{6.5cm}{0}{40}{40}{-132}{-60}
\caption{Comparison of template model ellipticals from different
theoretical codes for population synthesis.
Displayed are the predicted $M/L_{tot}$ ratio and $L_{PMS}/L_{MS}$ 
$V$-luminosity partition from the models by O'Connell (1976) [OC], Tinsley and Gunn (1976)
[TG], Pickles (1985) [PK], Guiderdoni and Rocca-Volmerange (1987) [GR],
Bruzual and Charlot (1993) [BC], Worthey (1994) [W], Bressan \etal (1994)
[BCF], and Buzzoni (1995) [B]. The effect of artificially doubling
Post-MS contribution (then relaxing FCT prescriptions) is displayed 
for Buzzoni's models by the dotted 
line.}
\end{figure}

A striking evidence from the figure is that model ellipticals by
Tinsley and Gunn (1976), Pickles (1985), Guiderdoni and Rocca-Volmerange (1987),
and Bruzual and Charlot (1993) do not fully comply with the SSP energetic 
constraint in the sense that all of them appear to be sensibly overestimating 
Post-MS contribution.

Focussing for example on the Bruzual and Charlot (1993) model, (Fig. 2, {\it left 
panel}) one can appreciate a noticeable scatter, of the order of $\pm 50$\%,
with respect to Buzzoni's (1995) calculations.
Note in addition that Post-MS excess in the model at older ages 
abruptly reverses for $t < 10$ Gyr. As far as cosmological predictions are 
concerned, diverging conclusions would be achieved, as 
shown in Fig. 2 {\it (right panel)}. In particular, while Bruzual and Charlot
would predict a redder apparent $B-V$ for low-redshift ellipticals 
with respect to Buzzoni's model, colors would then turn to be much bluer 
with increasing $z$.

\begin{figure}
{\plottwo {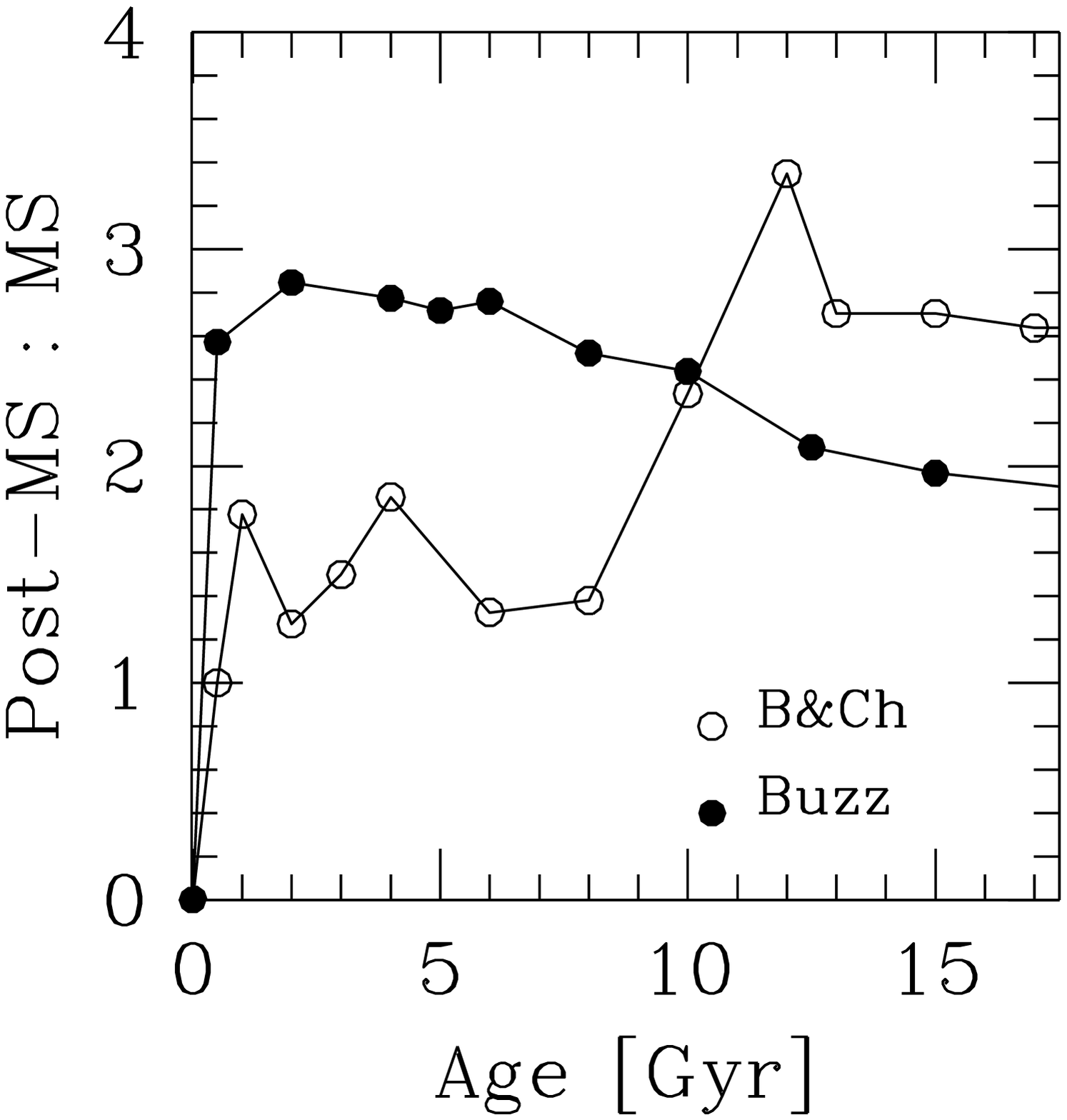} {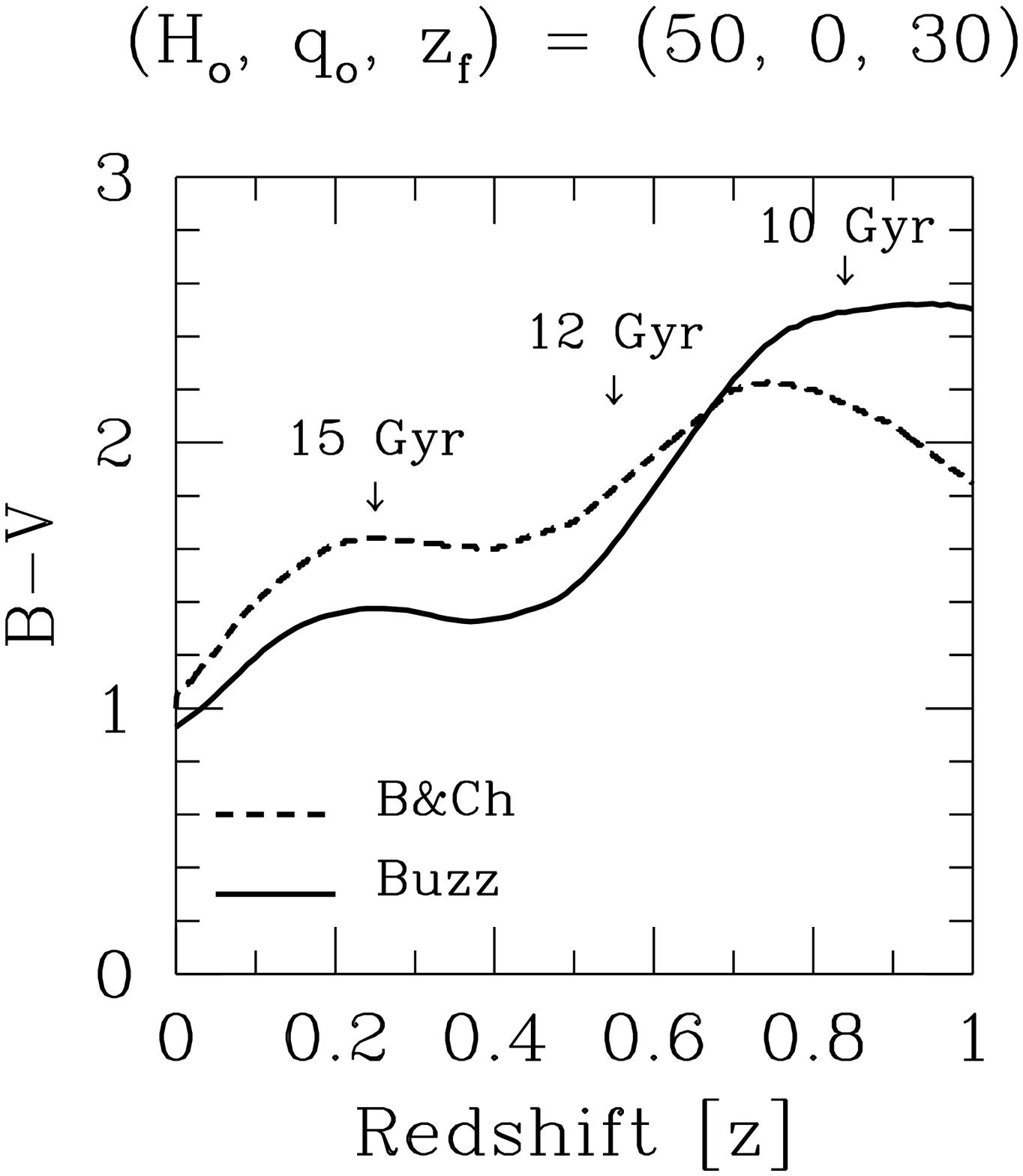}}

\caption{{\it (left panel)} - Bolometric evolution of template elliptical galaxies from Bruzual and Charlot 
(1993) ($\circ$) and Buzzoni (1995) ($\bullet$).
For a 15 Gyr SSP, $m_H$ in eq. (1) is of the order of 0.35 $M_\odot$ 
(Renzini and Buzzoni 1986) so that $L_{PMS}/L_{MS} \simeq 2$ in bolometric,
according to FCT prescriptions. \protect\\
{\it (right panel)} - Predicted apparent colors vs. redshift in a 
($H_o, q_o) = (50 km/sec/Mpc, 0)$ Universe. Absolute age of ellipticals 
is labeled along the curves assuming a redshift of galaxy formation
$z_f = 30$.}
\end{figure}

\section{Galaxy Back-in-time Evolution}

A new complete set of model galaxies of different morphological type,
spanning the whole Hubble sequence, has been produced via a semi-analytical
approach relying on SSP theory (Buzzoni 1997).

Each model consists of a spheroid component (that is a bulge+halo system)
and a disk. Appropriate luminosity partition among the different galaxy
subsystems with changing morphology, as well as adopted star formation 
history have been tuned by probing galaxy evolutionary status as observed
at present time (cf. Buzzoni 1997 for a full discussion of the theoretical 
details). A sample of our results is displayed in Fig. 3.\\
Two major features seem to characterize primeval galaxy evolution:

\begin{figure}

\plotfiddle{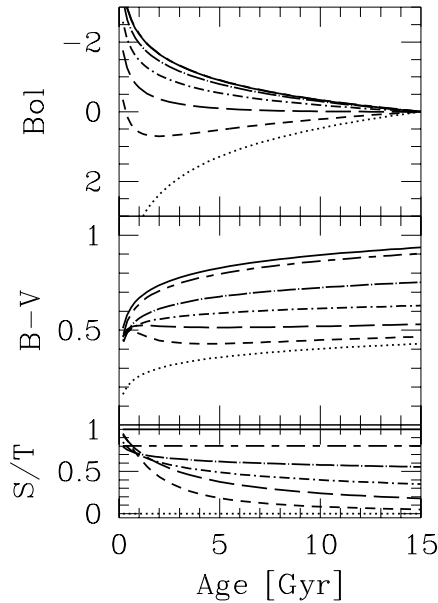}{7.7cm}{0}{50}{50}{-142}{-120}
\caption{Theoretical evolution of template model galaxies of different
morphological type according to Buzzoni (1997). The following types are 
considered in the three panels:
E (solid line), S0 (dot long-dashed), Sa (dot short-dashed), Sb (long-dashed),
Sc (short-dashed), Im (dotted). 
Upper panel reports bolometric evolution by normalizing model luminosity at
15 Gyr absolute age; middle panel shows the restframe $B-V$ color; in the lower 
panel, the galaxy morphological parameter $S/T$ measures the actual relative 
fraction of total bolometric luminosity provided by the spheroid (mostly bulge)
component. Every model (excepting Im galaxies) ends up to be bulge-dominated 
early in the past.}
\end{figure}

\begin{figure}
{\plottwob {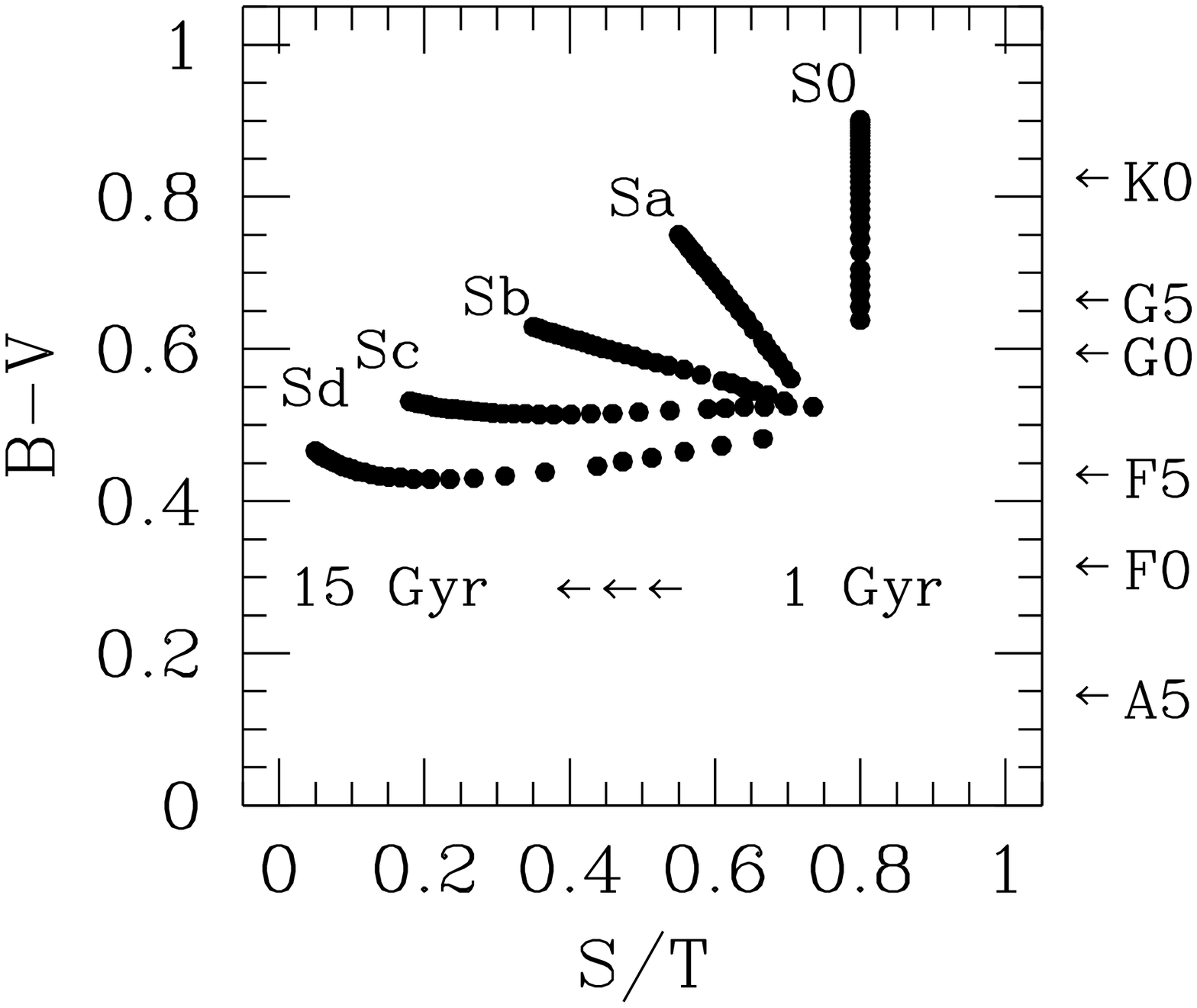} {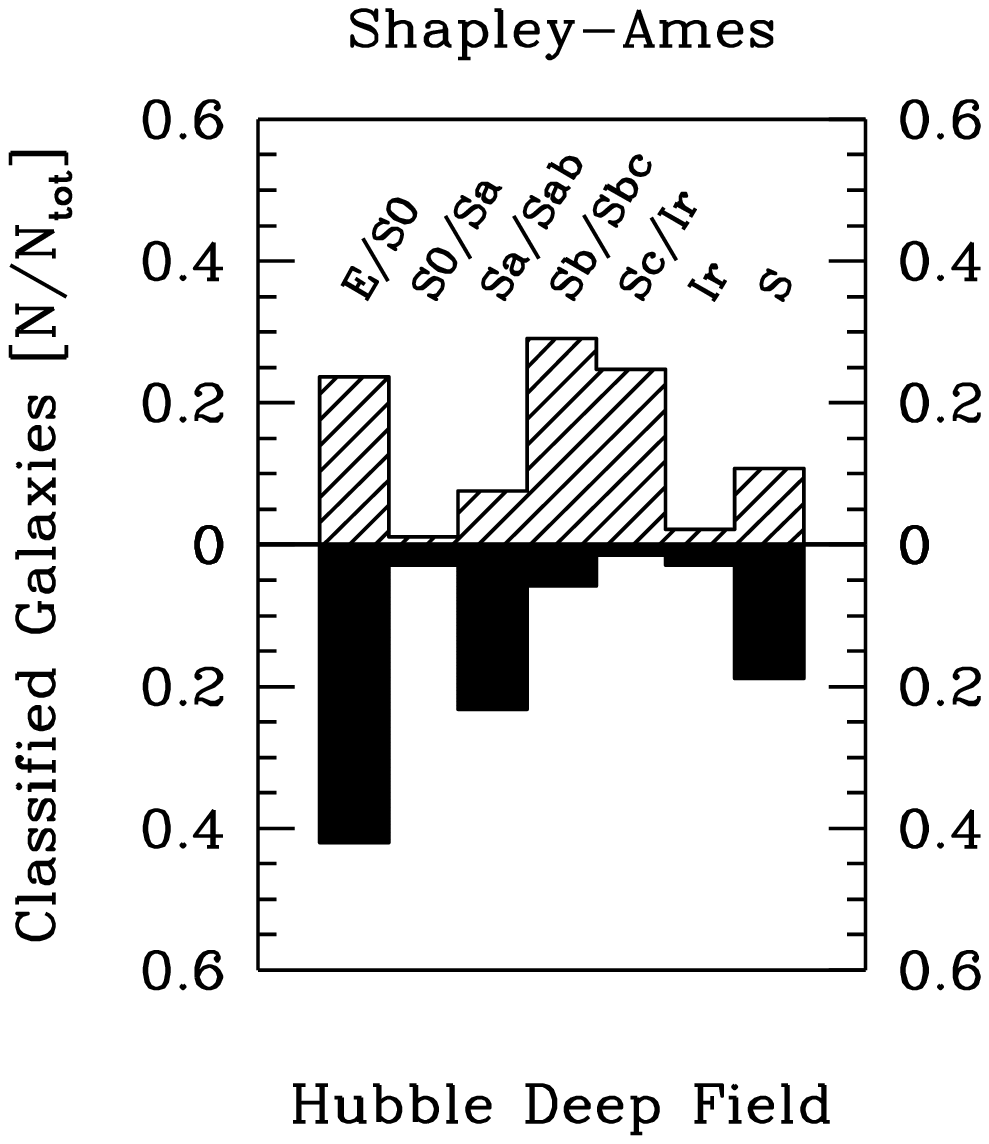}}
\caption{{\it (left panel)} - Restframe $B-V$ color for model galaxies
of different morphological type according to Buzzoni (1997). Evolution 
is computed for absolute age between 1 and 15 Gyr, as labeled.
Typical spectral types for stars of similar relevant color are marked
in the $y$-axis right scale.
The $S/T$ parameter refers to the actual bolometric contribution of
galaxy bulge. $S/T \to 1$ for very nucleated objects.\protect\\
{\it (right panel)} - Morphology distribution of galaxies in the HDF
compared with the local Shapley-Ames sample, according to van den Bergh
\etal (1996). A biased lack of Sb-Sc galaxies seems to occur in the
HDF in favor of E-Sa systems.}
\end{figure}

{\it i)} all systems currently displaying even a marginal 
luminosity contribution from the bulge (like for example in Sd spirals) have 
been largely dominated by the nuclear emission early in the past;

{\it ii)} for a wide range of galaxy star formation histories,
all morphological types tend to evolve back in time toward (restframe) bluer 
colors.

{\it The natural conclusion of our brief analysis is therefore that primeval 
galaxies would look like very nucleated blue objects.}

According to Fig. 4 {\it (left panel)}, there are at least two main
sources of bias that could induce (opposite) misleading interpretations 
of high-redshift data when neglecting evolutionary effects.

{\it i)} As far as {\it only} galaxy apparent colors are taken into account
(that is with no hints about morphology) one would be detecting a substantial 
deficiency of (intrinsically) red objects at high redshift, and an apparent
excess of blue (supposedly ``active'') galaxies. This might account for 
instance for Madau's (1997) results by matching HDF and the Canada-France 
redshift survey (Lilly \etal 1996).

{\it ii)} Conversely, if we {\it only} rely on the morphological features
of distant galaxies with no information about their colors, then one
would lead to conclude that early-type systems were dominating 
in the early Universe while spirals would ostensibly be 
missing at high redshift. An example in this sense is the HDF morphology
study by van den Bergh \etal (1996). As shown in Fig. 4 {\it (right panel)},
compared with low-redshift morphology distribution, in the HDF there is
a lack of Sb-Sc spirals in favor of an important excess of 
bulge-dominated systems (E and Sa spirals).


\begin{thebibliography}{99}  
\bibitem[]{}
Bressan, A., Chiosi, C., and Fagotto, F. 1994, {\it ApJS}, {\bf 94}, 63
\bibitem[]{}
Bruzual, G., and Charlot, S. 1993, {\it ApJ}, {\bf 405}, 538
\bibitem[]{}
Buzzoni, A. 1989, {\it ApJS}, {\bf 71}, 817
\bibitem[]{}
Buzzoni, A. 1995, {\it ApJS}, {\bf 98}, 69
\bibitem[]{}
Buzzoni, A. 1997, {\it ApJ} submitted\\
(see also {\sf http:$\backslash\backslash$www.merate.mi.astro.it$\backslash\sim$eps$\backslash$home.html})
\bibitem[]{}
Guiderdoni, B., and Rocca-Volmerange, B. 1987, {\it A\&A}, {\bf 186}, 1
\bibitem[]{}
Lanzetta, K.M., Yahil, A., and Fern\'andez-Soto, A. 1996, {\it Nature}, {\bf 381}, 759
\bibitem[]{}
Lilly, S.J., Le F\`evre, O., Hammer, F., and Crampton, D. 1996, {\it ApJL}, {\bf 460}, L1
\bibitem[]{}
Madau, P. 1997 in ``The Hubble Deep Field'', STScI Symp Series, eds. M. Livio, S.M. Fall and P. Madau (STScI: Baltimore) in press
\bibitem[]{}
Madau, P., Ferguson, H.C., Dickinson, M.E., Giavalisco, M., Steidel, C.C.
and Fruchter, A. 1996, {\it MNRAS}, {\bf 283}, 1388
\bibitem[]{}
O'Connell, R.W. 1976, {\it ApJ}, {\bf 206}, 370
\bibitem[]{}
Pickles, A.J. 1985, {\it ApJ}, {\bf 296}, 340
\bibitem[]{}
Renzini, A., and Buzzoni, A. 1986 in ``Spectral Evolution of Galaxies'', eds. C. Chiosi and A. Renzini (Dordrecht: Reidel) p. 195
\bibitem[]{}
Steidel, C.C., Giavalisco, M., Dickinson, M.E., and Adelberger, K.L. 1996, {\it AJ}, {\bf 112}, 352
\bibitem[]{}
Tinsley, B.M., and Gunn, J.E. 1976, {\it ApJ}, {\bf 203}, 52
\bibitem[]{}
van den Bergh, S., Abraham, R.G., Ellis, R.S., Nial, R.T., Santiago, B.X., and Glazebrook, K.G. 1996, {\it AJ}, {\bf 112}, 359
\bibitem[]{}
Williams, R.E. \etal 1996, {\it AJ}, {\bf 112}, 1335
\bibitem[]{}
Worthey, G. 1994, {\it ApJS}, {\bf 95}, 107
\end{thebibliography}
\end{document}